\documentclass[preprint]{aastex}

\slugcomment{}

\shortauthors{Terndrup et al.}
\shorttitle{Rotation of Low-mass Stars}

\newcommand{\kms}{km~s$^{-1}$}
\newcommand{\msun}{$M_\sun$}
\newcommand{\vsini}{$v \sin i$}
\newcommand{\Ha}{H$\alpha$}

\begin{document}

\title{Rotational velocities of low-mass stars in the Pleiades and Hyades}

\author{
Donald M. Terndrup\altaffilmark{1}, John R. Stauffer\altaffilmark{2},
Marc H. Pinsonneault\altaffilmark{1}, Alison Sills\altaffilmark{1},
Yongquan Yuan\altaffilmark{1},
Burton F. Jones\altaffilmark{3}, Debra Fischer\altaffilmark{4},
and Anita Krishnamurthi\altaffilmark{5}
}
\altaffiltext{1}{Department of Astronomy, The Ohio State University,
Columbus, OH 43210.} 
\altaffiltext{2}{Smithsonian Astrophysical Observatory, 60 Garden
Street, Cambridge, MA 02138.}
\altaffiltext{3}{Lick Observatory, Board of Studies in Astronomy and
Astrophysics, University of California, Santa Cruz, CA 95064.}
\altaffiltext{4}{Department of Astronomy, University of California,
Berkeley, CA  94720.}
\altaffiltext{5}{JILA, University of Colorado and National Institute of
Standards and Technology, Campus Box 440, Boulder, CO 80303.}

\begin{abstract}
We have obtained high-resolution spectra of 89 M dwarf members of the
Pleiades and Hyades and have derived radial velocities, \Ha\ equivalent
widths, and spectroscopic rotational velocities for these stars.
Typical masses of the newly-observed Pleiades and Hyades stars are
$\sim 0.4 M_{\sun}$ and $\sim 0.2 M_{\sun}$, respectively.  We combine
our new observations with previously published data to explore the
rotational evolution of young stars with $M \leq 0.4 M_\sun$.  The
average rotation rate in the Hyades (age 600 Myr) is about 0.4 that of
the Pleiades (110 Myr), and the mean equivalent widths of \Ha\ are also
lower.  As found in previous studies, the correlation between rotation
and chromospheric activity is identical in both clusters, implying that
the lower activity in the Hyades is a result of the lower rotation
rates.  We show that a simple scaling of the Pleiades rotational
distribution for $M \leq 0.4 M_{\sun}$, corrected for the effects of
structural evolution, matches that of the Hyades if the average angular
momentum loss from the Pleiades to the Hyades age is factor of $\approx
6$.  This suggests that the distribution of initial angular momenta and
disk-locking lifetimes for the lowest mass stars was similar in both
clusters.  We argue that this result provides further evidence for a
saturation of the angular momentum loss rate at high rotational
velocities.
\end{abstract}


\section{Introduction}

The rotational properties of stars, and how those properties evolve
with time, are a strong function of mass.  Somewhat less obviously, our
knowledge of those properties is also a function of mass because lower
mass stars are fainter and thus more difficult to observe.  For four
decades \citep{sle56,abt62}, we have known that B and A stars usually
have large rotational velocities, and that there there is
little evidence for angular momentum loss on the main sequence in this
mass range.  The mean rotational velocity decreases rapidly towards
lower mass within the F spectral class, and there is evidence for main
sequence spindown, particularly for late-F dwarfs \citep{kra67,ws97}.
The correspondence in the location of this rotational velocity break
with the appearance of the \ion{Ca}{2} HK lines in emission led
\citet{sku72} to associate both phenomena with the development of an
outer convective zone.  Non-radiative processes within the convection
zone lead to a temperature reversal above the photosphere and to a
steady-state wind which carries away angular momentum by coupling to
the stellar magnetic field, thus explaining the observed main sequence
spindown \citep[e.g.,][]{mac91}.

A good empirical understanding of the rotational velocity evolution of
main sequence G and K dwarfs has only been obtained during the past two
decades
\citep{sta84,sta85,sta89,sta91,sod93,all96,bou97,jon97,kri98,que98}.
These observations show that most stars in this mass range arrive on
the main sequence as relatively slow rotators, but that a significant
fraction (about 20-25\%) arrive on the main sequence as rapid rotators
with $20 < v \sin i < 200$ \kms.  Based on observations of open
clusters of different ages, the time scale for these rapid rotators to
lose most of their angular momentum ranges from tens to hundreds of
millions of years, with the time scale increasing with decreasing
stellar mass \citep[e.g.,][]{shl87,cha95}.  In order to be consistent
with these observations, theoretical evolutionary models require one or
more of the following features: core-envelope rotational decoupling
\citep{end81,pin90,all98}, coupling of the stellar rotation to that of
its circumstellar disk during pre-main sequence evolution
\citep{kon91,li93}, or saturation of the angular momentum loss rate
above a specified rotation rate \citep{kep95}.  The expectations from
theory will be discussed at greater length in $\S$ 4.

The most recent observational efforts have been to extend the available
spectroscopic data to lower $v \sin i$ limits \citep{del98,que98} and
to lower masses, approaching the limiting mass for hydrogen burning in
a few cases.  \citet{jon96} and \citet{opp97} have determined
rotational velocities for a small sample of stars in the the Pleiades
and Hyades with masses down to $\sim$0.1 \msun.  \citet{bas95} and
\citet{del98} have similarly obtained rotational velocities for very
low mass field stars.  In addition, there have been many studies of the
rotation periods of stars in young clusters as revealed from photometry
\citep[e.g.,][and references in these
papers]{all96,bou96,ch96,ps96,kh97,kh98,kri98,bar99,sta99,ter99}.  All
of these studies indicate that the time scale for angular momentum loss
is longer for lower mass stars, with stars near 0.1 \msun\ having
significant rotation rates even after stars at higher masses have all
spun down.  This facet of the observational database is best explained
by assuming that the critical rotation rate for saturation of the
angular momentum loss rate is a decreasing function of mass, a
conclusion supported by measures of chromospheric and coronal activity
in these stars \citep{ps96,bou97,kri98}.

In this paper, we provide new spectroscopic rotational velocities
(\vsini) for 89 M dwarfs in the Pleiades and Hyades.  When combined
with previously published data for these clusters, we can explore the
distribution of rotation rates in these two clusters much better than
in the past, particularly for low-mass stars (here $M \leq 0.4
M_\sun$).   We also analyze the strength of H$\alpha$ emission in the
two clusters, and discuss implications of our data for theoretical
modeling of angular momentum loss.

\section{Observations and Data Reduction}

\subsection{Target Selection and Observations}

Our new spectral data were obtained during three observing runs:  Nov.
19-21, 1996 on the KPNO 4m telescope\footnote{Kitt Peak National
Observatory, NOAO, is operated by the Association of Universities for
Research in Astronomy, Inc., under cooperative agreement with the
National Science Foundation.}, Dec. 4-5, 1996 on the Keck I
telescope\footnote{The W. M. Keck Observatory is operated as a
scientific partnership among the the California Institute of
Technology, the University of California and the National Aeronautics
and Space Administration.  The Observatory was made possible by the
generous financial support of the W. M. Keck Foundation.} using the
HIRES spectrograph \citep{vogt94}, and again on Dec. 4-5, 1997 with the
same setup.  At KPNO, we used the echelle spectrograph and the red long
camera, a slit width of 1.0 arcsec, a 31.6 lines/mm grating, and a
Tektronix 2048$^2$ CCD, resulting in an instrumental resolution of
about 37000 and wavelength coverage from 6000 to 8700 \AA.  With HIRES,
we used the standard echelle grating, a $0.86 \times 14$ arcsecond
slit, and a Tektronix $2048 \times 2048$ CCD with 24-micron pixels for
a spectral resolution of about 45000 and wavelength coverage from 6300
to 8700 \AA.  We used standard IRAF tools to do the basic CCD data
processing and also the order extraction, sky subtraction and
wavelength calibration.  At both telescopes, exposure times were
selected in order to obtain S/N $\sim 10$ near blaze center around 6600
\AA\ (for \Ha); given the redness of the program stars, this generally
resulted in S/N $\sim 15 - 20$ near blaze center for orders around 8500
\AA.

In the Pleiades, spectroscopic rotational velocities have previously
been obtained for a large, reasonably unbiased sample of stars down to
$V \sim 14$ \citep{sod93,que98}.  Below $V = 14$ ($M \simeq 0.6
M_{\sun}$), however, the fraction of known members with measured
rotational velocities decreases rapidly.  Even for the stars with
existing \vsini\ measures, most are from an ``early'' program (in the
1980's) with somewhat poorer spectral resolution resulting in \vsini\
limits of $9 - 10$ \kms\ for the slow rotators in the cluster
\citep[cf.][]{sta87}.  In order to obtain a better determination of the
distribution of rotational velocities for $M < 0.6 M_{\sun}$, we
devoted one of the KPNO 4m runs to obtain spectra of Pleiades members
with published $BVRI$ photometry in the magnitude range $14 < V < 17$.
We succeeded in obtaining spectra of 59 of these stars during the KPNO
run.   A small set of additional spectra for stars in this magnitude
range was obtained during a 1997 Keck HIRES run, bringing the total
number of Pleiads with new spectral data to 71.

In the Hyades, our goal was to constrain better the rotational velocity
distribution for (Cousins) $V - I_C> 3.0$.   Poor weather in Hawaii
during the time of our observing run significantly impaired the success
of our efforts, but we did manage to obtain spectra of 18 Hyades
members redder than $V - I_C = 2.9$.

Typical spectra in the region of H$\alpha$ and at $\sim$8400 \AA\ for
fast, moderate, and slowly rotating Pleiades members from the KPNO run
are shown in Figure 1, and a similar set of spectra for Hyades members
from the Keck observing run are shown in Figure 2.  Note the
systematic change in the H$\alpha$ profile as the rotational velocity
increases.

\subsection{Radial and Rotational Velocity Analysis}

Radial and rotational velocities for all of the program stars were
derived using the cross-correlation package developed by R.\ Hewett
and L.\ Hartmann, as described in \citet{har86}; see \citet{jon97} and
\citet{sta97a,sta97b} for other recent papers which have used this
package.  Briefly, a high signal-to-noise spectrum of a slowly rotating
star of similar spectral type to the program objects is chosen as a
template.  For a given spectral type, two or more orders of the echelle
spectra are chosen where there is a good set of intrinsically narrow,
moderate strength absorption features in the program objects, and where
there are few or no terrestrial (Earth atmosphere) features.  The
program star spectra for these echelle orders are then cross-correlated
against the template star spectrum, and the width and location of the
cross-correlation peak derived.  The calibration of this width to yield
an estimate of $v \sin i$ is determined by convolving the spectrum of
another high signal-to-noise slow rotator with rotational broadening
functions for a range of $v \sin i$ and then cross-correlating those
artificially broadened spectra versus the template star spectrum.

In all cases, we used high signal-to-noise spectra of \citet{gl69}
catalog M dwarfs as the calibrators.  For the KPNO spectra, we used
Gl~851 as the template star, and Gl~15B as the ``spin-up" star; for the
Keck spectra, we used spectra of Gl~905 and Gl~402 for these purposes.
These slow rotators have rotation rates which are considerably smaller
than our limit in this study:  Gl~15B has $v \sin i < 3.1$ \kms, Gl~402
has $v \sin i < 2.3$ \kms,  Gl~905 has $v \sin i < 1.2$
\kms\ \citep{del98}, and Gl~851 has $v \sin i = 2.5$
\kms\ \citep{mc92}.  To determine the rotational velocities of the slow
rotators amongst our program stars, we used echelle orders dominated by
TiO bands for the cross-correlation analysis; for rapid rotators, we
used echelle orders with several strong, relatively isolated atomic
absorption features.  We also cross-correlated a portion of an order
dominated by Earth-atmosphere O$_2$ lines to correct for drifts in the
spectrograph zero-point.

As a test of the radial velocity stability of the Keck HIRES, we
observed five M dwarf radial velocity standards from \citet{mar89}.
One of these (Gl~402) was used as our cross-correlation template star;
the mean shift of the other four relative to the tabulated Marcy \&
Benitz value was $-0.12$ \kms\ with a one sigma dispersion of only 0.3
km s$^{-1}$.  This shows that radial velocities accurate to better than
a half kilometer per second can be achieved with HIRES with no special
effort, at least for slow rotators (for example, we only used standard
spectral extraction routines and did not take ThAr spectra except at
the beginning and end of the night). The expected accuracy of the
derived radial and rotational velocities decreases with increasing
rotational velocity, with one sigma errors for both quantities being of
order 10\% of the \vsini\ based on our previous experience and on
comparison of our measurements for different echelle orders.  This
characteristic of the radial velocity errors can be readily seen in
Figure 3, which plots the radial velocity against \vsini\ for (upper
panel) our sample in the Pleiades and (lower panel) in the Hyades.  In
the Pleiades sample, which contains a number of rapidly rotating stars,
the scatter about the mean radial velocity is larger for the rapidly
rotating stars.   In addition to a higher scatter, there is some
indication that the radial velocities for the most rapidly rotating
stars are underestimated by 3--4 \kms.  That the scatter about the mean
at all velocities is within the expected errors (10\% of $v \sin i$)
leads us to conclude that few, if any, of our sample are not members of
these clusters as judged by their radial velocities.

The radial and rotational velocities for the Pleiades stars are
complied in Tables 1 (KPNO data) and 2 (Keck data), along with
\Ha\ equivalent widths (below) and cross references to each star in
various catalogs.  For most of these stars, the Cousins photometry
displayed in these two tables has been transformed from measured Kron
$V - I_K$ colors using the coefficients measured by \citep{bw87}.
Colors marked with a colon are transformed from lower-quality
photographic magnitudes tabulated by \citet{hhj93}.  Hyades data are
shown in Table 3.

Our radial velocities agree very well with previous determinations of
the mean velocities of much brighter stars in these clusters, which are
shown as dashed lines in the two panels of Figure 3.  For the Pleiades,
we find $\langle V_r\rangle = 6.0 \pm 0.3$ \kms\ (error of the mean),
in agreement with the value of $5.97 \pm 0.07$ \kms\ derived by
\citet{ng99} in their analysis of Pleiades stars with parallaxes from
Hipparcos.  For the Hyades, our new radial velocities average to
$\langle V_r\rangle = 39.7 \pm 0.4$ \kms, close to the value of $39.2
\pm 0.2$ \kms\ for Hipparcos-selected stars within 20 pc from the
Hyades cluster center \citep{per98}, and also in agreement with
$\langle V_r\rangle = 37.9 \pm 1.3$ \kms\ for non-binary dM stars in
that cluster \citep{sta97b}.  A better estimate of the accuracy of the
velocities in the Hyades can be made by comparing them to the
\citet{stef85} model of the dependence of velocity on position, using
their convergent point and Hyades space motion.  We find a scatter of
0.68 \kms, which reduces to 0.50 \kms\ if two stars with the largest
velocity differences (Re~119 and Re~158) are excluded.

\subsection{Additional Spectral Analysis}

For each of the program objects, we have measured the shape and
equivalent width of \Ha.  Two or three of the stars have essentially no
feature at \Ha; all of the rest have \Ha\ in emission.  For the slowest
rotators, the emission line profiles are steep-sided and boxy, with a
central reversal;  the profiles gradually change to centrally peaked
and approximately Gaussian in shape for the most rapid rotators, as
illustrated in Figures 1 and 2.  The equivalent widths, which are given
as positive values, range from barely detectable (0.1 \AA) to
moderately strong ($\sim 6$ \AA).  Two of the Pleiades dMe stars have
\Ha\ equivalent widths of $\sim 9$ \AA, but we suspect that their
steady-state chromospheric emission level is less than this and that
the equivalent width we have measured has been enhanced due to a
flare.

\citet{jon96} noted that two or three of the Pleiades dMe stars they
observed had broad \Ha\ wings in addition to the normal \Ha\ emission
profile seen in the other Pleiades stars.  They speculated that these
broad wings might be caused by mass motions produced during flares.  We
have reobserved two of those stars during the current program, and do
not find such wings on the new spectra.  The measured \Ha\ equivalent
widths for the new spectra are lower than for the previous spectra, in
accord with the idea that the broad wings may be associated with flares
(i.e., the stars were flaring during the previous observations). In
addition, we have found two new Pleiades dMe stars in our new set of
spectra which show broad \Ha\ wings similar to what had been found by
Jones et al.   The stars from the current program with broad \Ha\ wings
are HCG~143 (= T19B) and HGC~277 (= T105).  Both of these stars have
larger than average \Ha\ equivalent widths but fairly average
rotational velocities.

Finally, we have examined all of our spectra to check for the presence
of a detected lithium 6708 \AA\ absorption feature.  Because of the
relatively low S/N of our spectra, we generally would not detect
lithium absorption lines with equivalent width less than about 100
m\AA.  We do detect lithium in two of our program objects - HII 1756
($V = 14.1$, $V - I_C = 1.63$, $W_\lambda({\rm Li}) =  150$ m\AA) and
HCG~75 ($V = 14.45$, $V - I_C = 1.81$, $W_\lambda({\rm Li}) = 110$
m\AA).  Both stars are at the extreme blue end of our target list.
They are comparable in effective temperature to the latest type main
sequence stars in which lithium has previously been detected
\citep{gl94,jon96}, and have comparable equivalent widths to those
stars.  Both stars are slow rotators ($v \sin i \leq 7$ \kms and $v
\sin i = 8$ \kms, respectively).

\section{Analysis}

\subsection{Effective Temperatures}

We now proceed to adopt an effective temperature scale for the stars in
our sample.  The goal here is to combine our new data with previous
measures of $v \sin i$ in the Pleiades and Hyades ($\S$ 3.3, below).
The photometry for these stars is heterogeneous, with some having only
$B - V$ colors, others having only $V - I_C$, and a subset having both
colors.  We thus need temperature estimates on a common scale for both
colors.

Effective temperatures are derived from fitting a new generation of the
Yale Rotating Evolution Code (YREC) isochrones \citep{yrec92} to the
color-magnitude diagram of the Pleiades.  The new models, which include
the effects of rotation, are fully described by \citet{spt99};  they
incorporate many improvements from a number of groups to calculations
of low-temperature stellar atmospheres, opacities and equations of
state.  By including the physics of many molecules, \citet{ah95} have
created model atmospheres which are valid for low mass stars with
effective temperatures less than 4000 K.  The equation of state of
\citet{sau95} includes partial dissociation and ionization of hydrogen
and helium caused by both pressure and temperature effects, and is
applicable to both low mass stars and giant planets. Finally,
\citet{af94} added atomic line absorption, molecular line absorption
and some grain absorption and scattering to the usual sources of
continuous opacity to produce opacity tables which reach temperatures
as low as 700 K.  Since most previous atmospheres, opacities and
equations of state did not include the effects of molecules and grains,
these three improvements significantly increase our ability to model
the interior physics of very low mass stars \citep[also see][]{cb97}.

Isochrones were generated by computing evolutionary tracks at mass
intervals of $0.05 M_\odot$, and using linear interpolation between the
time points.  The zero-point for the age scale was set to the
deuterium-burning birthline as given by \citet{ps91}.  The manner in
which these isochrones were fit to the Pleiades data is illustrated in
Figure 4, which plots dereddened photometry for a compilation of stars
with previous $v \sin i$ measures (discussed in $\S$ 3.3 below) for
which $B - V$ photometry is available.  The photometry has been
dereddened using $E(B - V) = 0.04$, which is appropriate for most
regions of the Pleiades \citep{bre86,sta87}.  The dashed line in Figure
4 displays an isochrone for an age of 110 Myr, which was chosen as an
average between recent estimates using isochrones
\citep[e.g.,][]{mey93} and the somewhat higher values derived from the
detection of Li at low temperatures \citep{ssk98}.  (Our analysis is
insensitive to the exact choice of the Pleiades age, since the
mass-luminosity relationship is a very slow function of age near 110
Myr).  We adopted the metallicity of the Pleiades as solar;
\citet{boe90}, for example, derived [Fe/H] $= -0.034 \pm 0.024$ for
this cluster.

In generating the Pleiades isochrone, the \citet{ah95} model
atmospheres were used as a boundary condition (pressure at $T = T_{\rm
eff}$) on the interior models.  We chose to employ the empirical Yale
color calibration \citep{gdk87} to generate model colors instead of
relying on the color-temperature relations of \citet{ah95}.  The two
color calibrations have similar problems in matching the photometry.
Both predict colors which are far too blue for the lowest-mass stars,
though the match in $V - I_C$ is somewhat better in the \citet{ah95}
calibration, though not as good in $B - V$ for hotter stars.  We chose
to adjust the color-temperature calibration under the restricted
assumption that the errors were entirely in the color-temperature
calibration and not in the bolometric corrections \citep[cf.][]{jt98}.
The approach used here is discussed at greater length by \citet{pin00}.

The solid line in Figure 4 is the same isochrone adjusted to
match the photometry better, as follows:  For $(B - V)_0 \leq 0.75$,
the isochrone was used to determine the distance to the Pleiades using
a method discussed at length by \citet{pin98}.  In this method, the
distance modulus for each star is computed by differencing the
dereddened magnitude with the absolute magnitude of the isochrone.
This produces a distribution of distance moduli which is strongly
peaked at the mean distance to the cluster, but which has a tail toward
shorter distances for the binary stars.  The derived distance was taken
as the median distance modulus for stars with $(B - V)_0 \leq 0.75$ and
within 0.2 mag of this peak; for the Pleiades, this yields a distance
modulus of $(m - M_{\rm V})_0 = 5.6$.  For $(B - V)_0 \leq 0.75$, the
YREC isochrone fits the lower envelope of the Pleiades photometry quite
well, indicating that the model atmosphere colors and bolometric
corrections are accurate in this temperature range.

For $(B - V)_0 > 0.75$, we assumed that the YREC isochrone produced the
correct effective temperature and luminosity, and that the Yale 
bolometric correction was also correct in this temperature
range.  For cooler temperatures, we assumed that any differences
between the Pleiades photometry and the isochrone was caused by an
incorrect transformation between effective temperature and color;  thus
the adjustment process shifts a star of a given mass in color but does
not change its magnitude.  The isochrone was adjusted horizontally by
determining intervals of color which corresponded to mass intervals of
0.1 \msun, measuring the distribution of distance moduli for stars in
those color intervals with respect to the isochrone, then shifting the
isochrone to require that the median distance modulus be the same as
derived for the stars with $(B - V)_0 \leq 0.75$. (If the unadjusted
isochrone were too blue, for example, then it would lie below the
photometry, and the derived distances would be too low).  Down to $(B
- V)_0 = 1.5$, the required color shifts are small, everywhere less
than 0.065 mag.  We shall now refer to the adjusted YREC isochrone as
the ``empirical'' isochrone for the Pleiades.

In Figure 5, we display the YREC and empirical isochrones in $(V -
I)_0$ compared to available photometry (these are not always the same
stars as in Figure 4).  The symbols are the same as in Figure 4, except
we have added the $V - I$ photometry for the stars in this paper (open
triangles).  The production of the empirical isochrone in this color
(i.e., the determination of horizontal shifts to the unadjusted
isochrone to match the photometry) used the same method as in $B - V$.
A reddening of $E(V - I_C) = 0.06$ was used, computed from $E(B - V)$
using the relations in \citet{bb88}.  Here the color shifts are
significantly larger for stars at the faint end of the main sequence,
corresponding to $\Delta(V - I)_0 = 0.5$ mag at $M = 0.25$\msun.

The empirical isochrone was used to generate transformations between
$(B - V)_0$ or $(V - I)_0$ and effective temperature for stars in the
Pleiades.  These are shown in Figure 6, which shows the $B - V$
transformation in the lower panel and that in $V - I$ above.  The solid
line in each panel shows the color-temperature relation for the
empirical isochrone.  In the lower panel, the dashed line shows the
color-$T_{\rm eff}$ calibration discussed in \citet{sod93} for $(B -
V)_0 \leq 1.35$;  this is very slightly cooler than the empirical
isochrone.  In the upper panel, the dashed line shows the temperature
scale for K and M dwarfs employed by \citet{sta97a}, which is is based
on the \citet{bes79} color transformations for warmer stars and the
\citet{kir93} observations of M dwarfs.  The filled points are the
measured effective temperatures for individual M dwarfs in
\citet{kir93}.  The match between the temperature scale for the
empirical isochrone and that of Kirkpatrick et al.\ is very good for
the lowest masses ($T_{\rm eff} = 3200$ K, which corresponds to $M
\approx 0.1$\msun), even though -- as we saw in Figure 5 -- a rather
large adjustment in color was needed to match the Pleiades
photometry.    The open points in Figure 6 show temperatures derived by
\citet{leg96} for nearby M dwarfs;  their temperature scale is about
100 K cooler than that of \citet{kir93}.  As discussed by
\citet{sta98a}, the best effective temperature scale for the Pleiades
is probably intermediate between the two.  The temperatures in $V - I$
adopted here are considerably warmer (up to 100 K) at intermediate
temperatures than that in \citet{sta97a};  this is partly a consequence
of adopting an older age for the Pleiades than the value of 70 Myr used
in that paper, and partly a consequence of using different isochrones.

We computed effective temperatures for Pleiades stars independently
from $(B - V)_0$ and $(V - I)_0$, using the relations shown in Figure
6.  For stars with $(B - V)_0 > 1.35$, only the $(V - I)_0$ color was
used.  If both colors were available, the two temperatures were
averaged.  The use of the empirical isochrone assures that the
temperatures from both colors will be consistent with one another.  For
stars with photometry in both colors, the mean difference in
temperature between the two estimates was $\langle T_{\rm eff}(B - V) -
T_{\rm eff}(V - I)\rangle = - 27$ K (i.e., the scale from $(V - I)_0$
was slightly hotter), with a scatter of 110 K r.m.s.  Photometric
errors of about 0.03 in either color would produce this scatter in the
derived temperatures.

We also generated an isochrone appropriate for the Hyades, using the
new YREC models for [Fe/H] = $+0.13$ \citep{boe90}, an age of 600 Myr
\citep{per98}, and a distance modulus of 3.34 \citep{per98,pin98}, and
then applying the same shifts in color which were generated from
matching the photometry in the Pleiades.  The reddening adopted here
was $E(B - V) = 0.0$ \citep{cp66}.  The Hyades cannot itself be used to 
calibrate the color-temperature relation, because there is a significant
dispersion in distance for stars in the cluster.  \citet{pin98} compare
the YREC isochrones to Hyades stars with Hipparcos parallaxes, which
are all much brighter than the stars in our sample.  The match
between the isochrones and the photometry is very good.

\subsection{Mass estimates}

Because the Pleiades and Hyades are not coeval and have different
metallicities, the relationship between effective temperature and
stellar mass is not identical for the two clusters.  Consequently, a
comparison of the rotation rates for the two data sets is sometimes
better made in mass rather than effective temperature or color.

We derived mass estimates for the stars in our sample using polynomial
fits between mass and $V - I_C$ for the Pleiades and Hyades empirical
isochrones described previously.  The aim here was to adopt a
color-temperature relation (rather than a luminosity-mass relation) so
that binary stars would not be assigned higher masses than single stars
of the same observed color.  For $M < 0.4 M_\sun$, the Hyades stars
are, on average, about 90 K cooler at each mass than stars in the
Pleiades.  On this scale, the typical mass of the newly observed
Pleiades stars is 0.4 $M_\sun$, and is 0.2 $M_\sun$ in the Hyades.

For the faintest, reddest stars in both clusters, the method used to
estimate masses based on effective temperature produces masses which
are systematically too low (i.e., substellar) compared to other
estimates in the literature for such faint stars.  We have therefore
considered an alternative procedure which relies more on a star's
luminosity rather than its inferred temperature.  For this alternative
method, we adopte the \citet{mon92} estimate of the $I$-band bolometric
correction.  Using the available I$_C$ photometry, we then computed the
bolometric magnitude for each star.  Then we derived the mass by
interpolation using the empirical isochrones for the two clusters.  For
$M \geq 0.25 M_\sun$, the two methods gave the save average relation
between color and mass, though the mass estimates could differ
considerably for probable binaries, where the latter method yields a
higher mass because of the brighter $I_C$ magnitude.  Below $M = 0.25
M_\sun$, the alternative method yielded higher masses, with the coolest
stars in both clusters having mass estimates in the range 0.12--0.15
$M_\sun$, in accord with masses estimated for these stars in previous
papers \citep{sta94,sta95}.  In order to avoid assigning substellar
masses for the reddest stars, we adopted a compromise mass scale,
taking the masses from the empirical isochrones for $M \geq 0.25
M_\sun$, and smoothly interpolating to the alternative mass scale for
lower masses.  As we are mainly concerned with broad trends in the
rotational properties with stellar mass, , in particular by discussing
the stars in intervals of mass (below), none of the details of our
analysis depend sensitively on the adopted mass scale.

\subsection{Distribution of rotational velocities}

In Figure 7, we plot the $v \sin i$ data for the Pleiades (upper panel)
and the Hyades (lower panel) against effective temperature.  The solid
points in this figure are for the stars in this paper, while open
points are for additional data in the literature for the Pleiades
\citep{bas95,que98} and for the Hyades \citep{sta97b}.  In the Hyades,
no additional rotation rates from the literature are plotted for
$\log(T_{\rm eff}) > 3.6$, which corresponds to $V - I_C < 1.9$;  in
this temperature range, \citet{z93} detected no stars with $v \sin i >
10$ \kms\ (the limiting velocity in that study) up to $T \approx 6000$
K, except for some tidally locked binaries \citep[see also][]{tho93}.
Studies of the rotation periods of Hyades stars earlier than spectral
type K7 also show that all are rotating slowly \citep{rad87}.  At the
bottom of each panel of Figure 7, we have also indicated the stellar
masses from the Pleiades and Hyades empirical isochrone, derived as
described previously.

Our new data increases the number of stars with spectroscopic measures
of rotation for $M < 0.5 M_\sun$.  Our goals here are twofold:  to
compare the rotational distribution of low-mass stars to the
well-studied sample of stars near solar mass, and to explore the
characteristics of angular momentum loss at the bottom of the main
sequence.  As discussed more extensively in $\S$ 4, the range of
rotation rates present at each mass sets strong constraints on the
mechanisms for angular momentum loss on the main sequence and earlier.

The distribution of rotation rates in the Pleiades has a strong
dependence on mass.  For $M > 0.9 M_\sun$, most stars fall near a lower
envelope which is a decreasing function of mass, going from $\approx
30$ \kms\ at $M = 1.25 M_\sun$ to $\approx 8$ \kms\ from 0.6 -- 0.9
$M_\sun$.  At higher masses there is a wide distribution of rotation
rates from $v \sin i \leq 7$ \kms\ up to $v \sin i \approx 100$ \kms, a
factor of at least 15 in angular momentum per unit mass.  Most of the
stars in this mass range are slowly rotating, with $v \sin i \sim 7$
\kms, but there is also a small fraction of stars with quite rapid
rotation ($v \sin i \approx 80$ \kms), producing an apparently bimodal
distribution of rotation rates.  The range of rotation rates declines
to lower masses:  below about 0.3 \msun, there are few or no stars with
upper limits on $v \sin i$ (typically 7 \kms), and the upper envelope
of the rotation rates is decreasing with decreasing stellar mass.  It
would also seem that the distribution of rotation rates changes 0.5
\msun, in that the distribution is uniform and not bimodal, as found at
higher masses \citep[cf.][]{jon96}.  This flat distribution is quite
unlike the distribution expected for a narrow range of rotation rates
broadened by random inclination angle, which would be peaked toward
stars with high apparent rotation.  The full range of rotation rates
near 0.2 \msun\ is only about a factor of 3--4.  The fall in the upper
envelope of rotation rates corresponds to a fairly constant maximum
{\it angular} rotation rate, since the stellar radius declines with
decreasing mass \citep[this is shown in Fig.\ 6 of][]{jon96}.

In Figure 8, we present histograms of the distribution rotational
rates, arranged in three groups from highest to lowest masses (top to
bottom).  The histograms show the number of stars per intervals of
$\Delta\log_{10}(v \sin i) = 0.151$ (i.e., intervals of $\surd 2$ in $v
\sin i$).  Stars with limits on $v \sin i$ are plotted in the bin
corresponding to the limiting value;  the bins in each histogram were
constructed so that the stars which have $v \sin i \leq 7$ \kms\ in
this study would be plotted in the leftmost bin.  The mass ranges used
in each histogram are shown;  the upper limit for the highest mass bin
was set to exclude stars in that part of the $v \sin i$ distribution
where the average velocities are increasing with higher temperature.
In addition, the distribution of velocities for our combined Hyades
sample is shown as a dashed histogram in the lowest panel of the
Figure.

Although there are some Pleiades stars in all bins of $v \sin i$
for $M \geq 0.4$ \msun, the separation between the majority
slow-rotator group and the fast-rotators is apparent in the upper two
panels of Figure 8.  In these mass bins, the fast rotators have a peak
in the distribution of velocities near $v \sin i \approx 80$ \kms.  The
peak of velocities for the slow rotators is only a few \kms\ for the
hottest stars shown;  it possibly rises to about $7-9$ \kms\ for the
interval $0.4 < (M / M_\sun) < 0.6$, though this is less certain
because many of the slowest rotators in this mass range have limits on
$v \sin i$.

The data in the lower panel of Figure 8 show that the lowest mass stars
have a flat distribution of rotational velocities.  There is no
evidence for the bimodal distribution observed for higher mass stars.
The distribution of velocities is remarkable also in that there are few
stars rotating at or below the observational limit in this study.   A
similar lack of slow rotators has been seen in other young clusters
\citep{all96}.  The Hyades still has a number of stars with limits on
the velocity, at least down to $M \sim 0.2$ \msun.  Because of the
number of stars with limits in the Hyades, it is not possible to
estimate accurately if the range of velocities is much larger than the
factor of 3--4 in the Pleiades, though it is certainly not as small as
would be produced by a very narrow rotation distribution viewed at
random inclination angles.

\subsection{H$\alpha$ emission}

Figure 9 shows the correlation between the equivalent width of \Ha\ and
stellar temperature.  Stars in the Pleiades are shown as open symbols,
while those in the Hyades are displayed as filled symbols.  The
Pleiades data come from this study and from \citet{jon96}.  The Hyades
points come from this paper, \citet{sta97b} and \citet{jon96}.  The
different symbols indicate the source of the data, as outlined in the
figure caption.  In the case where we reobserved stars from earlier
work, only the \Ha\ value from this paper is plotted.

In both clusters, the well known trend towards higher \Ha\ widths at
lower temperatures is readily seen.  The distribution of \Ha\ for the
Hyades hardly overlaps at all with that of the Pleiades:  the maximum
equivalent width at each temperature for the Hyades is about equal to
the minimum in the Pleiades.  Thus we see a decline in chromospheric
activity with cluster age \citep[cf.][]{cal96}.  Note that the strong
rise of \Ha\ equivalent widths towards cooler temperatures is mainly an
effect of the rapidly falling continuum near the \Ha\ line; the stars
actually have a falling luminosity in \Ha\ with temperature
\citep[cf.][]{you89,bas95}, consistent with the picture sketched above
that the rate of angular momentum loss is declining with decreasing
stellar mass.

Figure 10 shows the correlation between $v \sin i$ and \Ha\ for the
Pleiades and Hyades stars plotted in Figure 9.  The masses have been
derived from the empirical isochrone, and the stars have been divided
so that each bin contains approximately the same number of stars.   The
scale for all three panels is identical.  At all temperatures, the
\Ha\ rises only slowly with rotation, except for slow rotation rates
where the rise with rotation is much steeper.  Such a change in slope
is more evident when the ratio of rotation speed to the convective
overturn time is considered \citep{ps96,kri98}, and has been used to
argue that the magnetic field strength and angular momentum loss rate
saturate at higher rotation rates.  In the bottom panel of Figure 10,
which shows the data for the lowest temperatures, there is a
significant number of both Hyades and Pleiades stars.  There are some
stars in both clusters with \Ha\ width considerably higher than the
mean value, which may represent stars which were flaring at the time of
observation.  With the exception of these points, the correlation
between rotation and chromospheric activity is identical for both
clusters.  This means that the average \Ha\ strength and the average
rotation rate decline with age by the same factor.  It is thus not
correct to use the \Ha\ strengths as a direct indicator of age as
occasionally has been proposed in the literature;  we see here that
stars of a factor of 5.5 different in age are indistinguishable if they
have the same rotation rates.  It is still true, however, that one can
use the distribution of \Ha\ vs.\ color as an age indicator, since this
relies on the mass dependence of spindown times as well as the
chromospheric response \citep[cf.][]{sta97a}.

\section{Discussion:  The evolution of rotation rates at low mass}

While our new data do not push the study of rotation into new territory
-- rapid rotation at low stellar masses has been seen before -- they do
provide a larger statistical sample at low masses.  The most important
aspect of the data is that we have actual measures of $v \sin i$,
rather than just limits, for most low-mass stars in the Hyades.  This
provides us with the chance to explore the nature of and time scale for
angular momentum loss over the age range sampled by the two clusters
($\approx 110$ to 600 Myr).

In particular, we pose this question:  can we expect that the
rotational distribution of low mass stars in the Pleiades will evolve
to look like that of the Hyades when the former cluster is a factor of
$\approx 5.5$ older?  If we can, then we provide independent support
for the current paradigm employed in theoretical models of angular
momentum evolution \citep[][and references therein]{kri97}.   If not,
then we may be uncovering evidence that the conditions of early
stellar evolution (initial angular momentum or lifetimes of stellar
disks) differed in the two clusters, or that the current evolutionary
paradigm is inadequate.  \citet{spt99} provide a direct comparison
between theory and the available data in these and other young
clusters.

There are four principal ingredients of the current theoretical
paradigm:  (1) Young stars begin their lives as fully convective
objects; because the time scale for convective transport is much
smaller than that for angular momentum loss, the stars rotate as solid
bodies\footnote{What matters for the internal transport of angular
momentum is the radial dependence of rotation.  The models we
are using do not handle latitudinal circulation.  In the Sun,
the convection zone shows latitudial differential rotation, but
this is nearly constant with radial depth \citep[e.g.,][]{tho96}.
} and their angular momentum profile with depth is fully specified
by the surface rotation rate, which is typically modeled to be like
that of T Tauri stars \citep{ch96}.  (2) For some time the surface
rotation periods of the stars, now contracting toward the main
sequence, are held nearly constant by magnetic fields threading between
the star and its circumstellar disk \citep{kon91,cam93,ccq95,kep95}.
When the disk and star decouple, the surface rotation rate rises as the
star continues its contraction.  (3) Angular momentum loss continues
via a magnetized wind during the evolution to the main sequence and
well into the early main sequence lifetime.  In a linear dynamo the
angular momentum loss rate is proportional to the cube of the surface
angular velocity \citep{wd67,kaw88}.  The resulting spindown for the
rapid rotators would, however, take place on such a short time scale as
to be in conflict with the maximum rotation rates seen in young
clusters \citep{pin90}.  Observational data on the correlation between
chromospheric and coronal activity suggest that there exists a
saturation threshold which reduces the angular momentum loss rate at
rapid rotation \citep{noy84,ps96,kri98}.  (4) At any period in the
evolution, the surface rotation rates can also depend on the mechanism
of and time scale for internal angular momentum transport.  For stars
with significant radiative cores, this may result in a core and surface
which spin at different rates.  Since the size of the radiative core
depends on the stellar mass and the location of the star along an
evolutionary track, there are stellar masses and cluster ages where the
differences between solid-body and differentially rotating models are
more pronounced \citep{kri97}.

The existence of the fast rotators at higher masses requires both that
some stars are tied to their disks for only a short time ($\leq 1$ Myr)
and that the rotational spindown is saturated above some threshold.
This latter property is usually modeled as 
\begin{eqnarray}
{{dJ} \over {dt}} 
 & =  -K_\omega \omega \omega_{\rm crit}^2, &  
   \omega > \omega_{\rm crit}  \\
 & =  -K_\omega \omega^3, &  
   \omega \leq \omega_{\rm crit}  
\end{eqnarray}
where $J$ is the angular momentum, $\omega$ the angular rotation rate,
and $\omega_{\rm crit}$ is the angular rotation rate where saturation
occurs.  The observation that the upper envelope of the rotational
distribution rises to lower masses for $M \geq 0.6 M_\sun$ (Fig.\ 7) is
typically modeled by having the saturation threshold be a function of
mass \citep[e.g.,][]{kri97}, with saturation occurring at slower
angular velocities in lower-mass stars, which then experience less
angular momentum loss early on \citep{bar96}.

Modeling the rotation rates in young clusters is not free of
ambiguities because the various ingredients of the current paradigm are
not independent.  In particular, the production of a dispersion of
rotation rates at a given mass could be the result of a dispersion in
initial angular momentum or in disk-locking lifetimes.  The normal
approach \citep{kri97,spt99} is to start all stars with a fixed initial
angular spin rate, set by observations of T Tauri stars, and
subsequently to produce a dispersion by a distribution of disk
lifetimes.  There is a degeneracy between the derived $\omega_{\rm
crit}$ and the assumed initial rotation rates at the stellar
birthline:  if very young stars have higher rotation rates than those
usually taken, as indicated by the recent observations in Orion by
\citet{sta99}, then higher values of $\omega_{\rm crit}$ are required.
(The Stassun et al.\ paper also demonstrates that there exists a wide
range of rotation periods even at an age of 1 Myr.)

Here we can ignore the issue of what produces the dispersion in
rotation rates at low masses, instead focusing on how this dispersion
would change with time.  The models of \citet{spt99} predict that stars
with $M \leq 0.4 M_\sun$ should still be rotating with $\omega \geq
\omega_{\rm crit}$ even at the age of the Hyades,  because without
saturation of the loss law all the stars would be rotating at lower
rates than is observed.  Now the angular momentum can be written as $J
= I\omega$, where $I$ is the momentum of inertia.  If $I$ is a constant
or varies much more slowly with time than $\omega$, then in the
saturated case we have
\begin{equation}
{1 \over \omega}{{d\omega} \over {dt}} 
 = -{{K_\omega} \over {I}} \omega^2_{\rm crit},
\end{equation}
which has the solution $\omega(t) = \omega_0 e^{-at}$, where $a$ is a
constant, taking $\omega \rightarrow 0$ as $t \rightarrow \infty$.  Thus
once $\omega_{\rm crit}$ is set (though it can vary with mass), all
stars of a given mass will spin down by the same factor over an
interval of time, and the ratio of $v \sin i$ between the fastest and
slowest spinners will be constant.  Alternatively, if all stars were
spinning with $\omega < \omega_{\rm crit}$, we would have
\begin{equation}
{{d\omega} \over {dt}} = 
-{{K_\omega} \over {I}} \omega^3,
\end{equation}
which has the solution $\omega(t) \propto t^{-1/2}$.  With the passage
of time, the rapid rotators would experience greater loss, and the
ratio of $v \sin i$ between the fastest and slowest spinners will
decline.  

The data in Figures 7 and 8 show that the range of rotation at the
lowest masses is similar in both clusters (specifically, we concluded
that the range in rotation in the Hyades is at least as large as in the
Pleiades). This supports the idea that the appropriate value of
$\omega_{\rm crit}$ for these stars has to be sufficiently low that
even the slow spinners in the Hyades still have $\omega \geq
\omega_{\rm crit}$.  In this case, we should be able to find a constant
$q$ (where $0 \leq q \leq 1$) to scale the Pleiades velocity
distribution from the observed $N(v \sin i)d(v \sin i)$ so that $N(qv
\sin i)d(v \sin i)$ matches the observed distribution in the Hyades
(here we mean that the shape of the velocity distributions would be
identical).  The value of $q$ would then calibrate the constant
$K_\omega$ in equations 1 and 2.  This turns out not to be possible
without correcting for fact that the lowest mass stars in the Pleiades
are still contracting towards the main sequence.  For example (using
the isochrones discussed in $\S$ 3.1), the radii of Hyades stars with
$M = 0.2 M_\sun$ are 0.8 that of the Pleiades, and the mass where
Hyades stars are on the main sequence is close to $0.5 M_\sun$.  Were
there no angular momentum loss, the low-mass Pleiads would spin up as
they contracted, and would have higher rotation rates at the age of the
Hyades.  That the rates in the Hyades are, in fact, about a factor of
two lower than in the Pleiades indicates that the angular momentum loss
is larger than any increase in rotation rates during the contraction.

We modeled the future evolution of the Pleiades stars as follows.  At
each mass $M$, let $r_i(M)$ be the radius at the age of the Pleiades,
and $r_f(M)$ be the radius at the age of the Hyades.  If the contraction
towards the main sequence is homologous, then the moment of inertia $I$
would scale as $r^2$.  Conservation of angular momentum requires
$I\omega =$ const., and so we should expect the rotation rates to
increase by a factor of $(r_i / r_f)^2$ in the absence of loss.  Then,
if as argued above, there should be a constant factor $q$ reduction in
the observed rotation rates from angular momentum loss, a Pleiades star
currently observing at a rate of $v_i = v \sin i$ will be have $v \sin
i = q v_i (r_i / r_f)^2$ at the age of the Hyades.

We calculated an expected (future) distribution for the Pleiades by
computing the factors $(r_i / r_f)^2$ for each star, using the masses
derived from the empirical isochrones.  Then the resulting rotation
rates were reduced by a factor $q$, but were set to a minimum value of
$v \sin i = 6$ \kms\ to match the limits in our current study. For each
$q$, we used a K-S test to compare the cumulative distribution $F(v\sin
i$) of scaled Pleiades velocities to that in the Hyades.  Values
outside the range $0.10 \leq q \leq 0.23$ could be excluded at the 95\%
confidence level; the best match was achieved for $q = 0.17$, for which
the probability that the two distributions could be drawn from the same
sample was $P_{\rm KS} = 0.90$.  We illustrate the good match which
results from this computation in Figure 11, where thick line shows the
observed cumulative distribution in the Hyades, while the thin line
shows the distribution of scaled velocities in the Pleiades.

This exercise sets a average time scale for the angular momentum loss
below $m = 0.4 M_\sun$.  The factor of six loss in the angular momentum
between the age of the Pleiades and that of the Hyades means that the
e-folding time scale for loss is near 250 Myr for $M \leq 0.4 M_\sun$
(the range of $f$ excluded by the K-S test corresponds to an estimated
average time scale of $245 \pm 55$ Myr).  This would imply that the
fastest rotators in the Hyades, now spinning near $v \sin i = 30$ \kms,
would have undetectable rotation rates after 1-2 billion years.

The similarity in the shapes of the two distributions in Figure 11
shows that the distribution of initial rotation rates for low-mass
stars in the two clusters must not have been very different, and
provides a verification of the existence of saturation in the angular
momentum loss at high rotation rates.  This also explains the change in
the distribution of rotation rates near $M = 0.5 M_\sun$:  at higher
masses, the rotation rates fall below $\omega_{\rm crit}$ by the age of
the Hyades, if they are not already so at the age of the Pleiades.  We
therefore see a reduction in the range of rotation rates by the age of
the Hyades, which would not be the case if all stars had $\omega >
\omega{\rm crit}$.

At present, only a few young clusters have sufficiently large samples
at these low masses \citep{spt99} for comparison to the Pleiades and
Hyades.  Data in other clusters could potentially show that the
distribution of initial rotation rates was universal or that it depends
on cluster mass, stellar density, binary fraction, or other factors.  A
more detailed comparison of the current data with theoretical models of
angular momentum evolution might allow a better calibration of
$\omega_{\rm crit}$ for low-mass stars.  This may remove some of the
current ambiguities in the modeling, specifically between the
disk-locking lifetime and the initial rotation rates.

\acknowledgments 
JRS acknowledges support from NASA Grants NAGW-2698 and NAGW-3690.  DT
and MHP acknowledge support from NASA Grants NAG5-7150 and NSF grant
AST-9731621.  AS acknowledges support from NSERC.  We are grateful
to the referee, who made many valuable suggestions.

\clearpage



\figcaption[]{
Examples of spectra at H$\alpha$ (left panels) and near 8400
\AA\ (right) for stars in the Pleiades, ordered from top to bottom by
decreasing $v \sin i$.}


\figcaption[]{Same as Figure 1, but for three stars in the Hyades.}


\figcaption[]{
Radial velocities of M stars in the Pleiades (upper panel) and the
Hyades (lower panel).  Open circles denote stars with measured $v \sin
i$, while triangles indicate upper limits.  The dashed line in each
panel shows the mean velocity of large samples of stars in the Pleiades
\citep{ng99} and in the Hyades \citep{per98}.  As discussed in the
text, errors in each point are better than 1 \kms\ for slowly rotating
stars, but are about 10\% of the $v \sin i$ value for rapidly rotating
stars.}


\figcaption[]{
Color-magnitude diagram for the Pleiades and comparison to model
isochrones.   The photometry has been dereddened using $E(B - V) =
0.04$.  The long dashed line shows the YREC models for an age of 110
Gyr and using the Yale color calibration \citep{gdk87}, 
shifted to a distance modulus of 5.6.  The solid line shows the
empirical isochrone for the Pleiades, derived as discussed in the
text.  The YREC models were adjusted by altering the color-temperature
relation so that they matched the empirical isochrone. 
}


\figcaption[]{
Same as Figure 4, but in $(V - I)_0$. Open circles show photometry in
the Pleiades which have previous measures of $v \sin i$,
while the triangles are photometry for the stars observed in this
paper.
}


\figcaption[]{
Comparison of color-temperature scales in (upper panel) $V - I_C$ and
(lower panel) $B - V$.  The solid lines show the effective temperature
at each color from the empirical isochrone for the Pleiades.  The
dashed line in the lower panel is from \citet{sod93}, while that in the
upper panel is that employed by \citet{sta97b}, based on data in
\citet{kir93} and color transformations by \citet{bes79}.  The solid
points in $V - I$ are for some individual M dwarfs in \citet{kir93},
while the open points with error bars are for stars in \citet{leg96}.
}


\figcaption[]{
Data on rotation rates in (top) the Pleiades and (bottom) the Hyades.
Filled points are from this study, while open points are from other
papers as described in the text.  Limits are shown as downward pointing
triangles.  The numbers at the bottom of each panel show the
equivalent masses from the empirical isochrone in solar units, derived
as discussed in the text.}


\figcaption[]{
Distribution of rotation rates for three intervals of effective
temperature.  The solid line shows the number of stars in equal
intervals of $\log_{10}v \sin i$ for the Pleiades.  The dashed
line in the lower panel shows the distribution of rotation rates
for the low-mass stars in the Hyades.
}


\figcaption[]{
Width of H$\alpha$ emission plotted against color.  Filled symbols are
for the Hyades sample in this paper plus a few points from Jones et
al.\ (1996);  these are shown as upward- and downward-pointing
triangles, respectively.  Open circles are for Pleiades stars in this
paper, while crossed open circles are from Jones et al.\ (1996).
}


\figcaption[]{
Correlation between rotation and H$\alpha$ width.  Symbols are the same
as in Figure 9.  The data have been divided into mass bins using the
mass estimates derived as described in the text.  Points with upper
limits to $v \sin i$ are plotted at that limit.
}


\figcaption[]{
Cumulative distribution $F(v\sin i)$ of rotation rates in the Hyades
(thick line) compared to that for the Pleiades at an age of 600 Myr
(thin line), computed from the observed velocities as described in the
text.
}

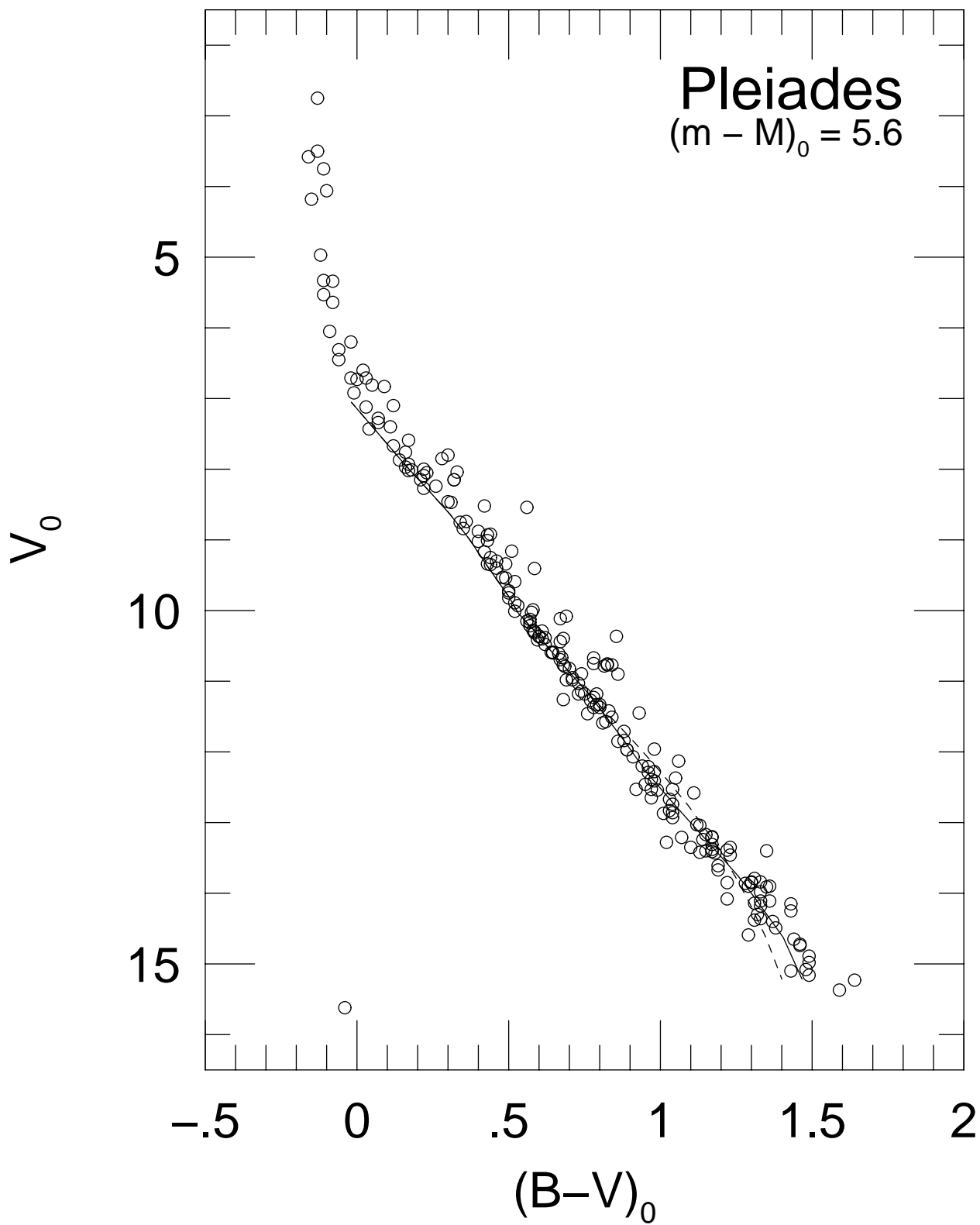
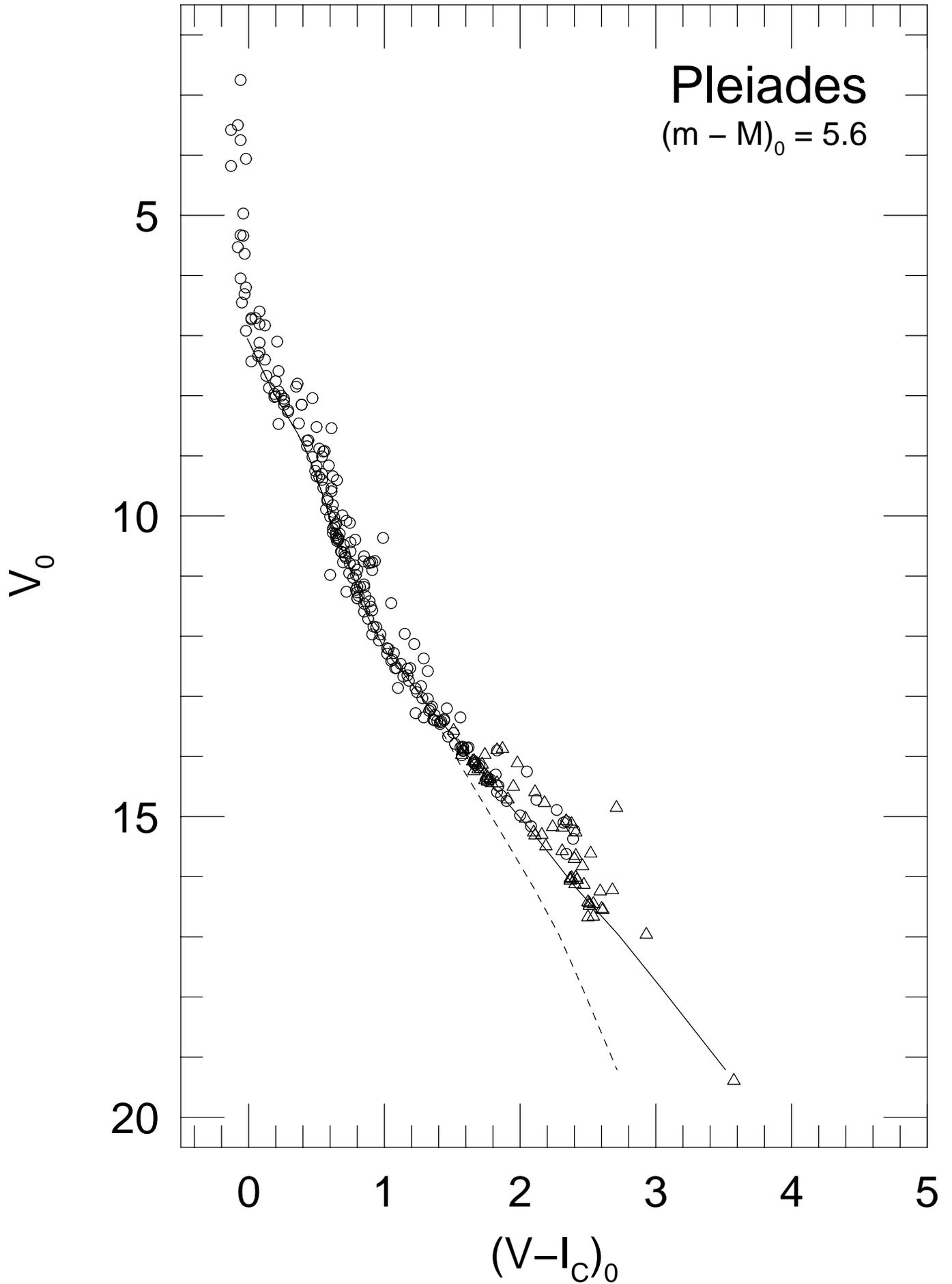
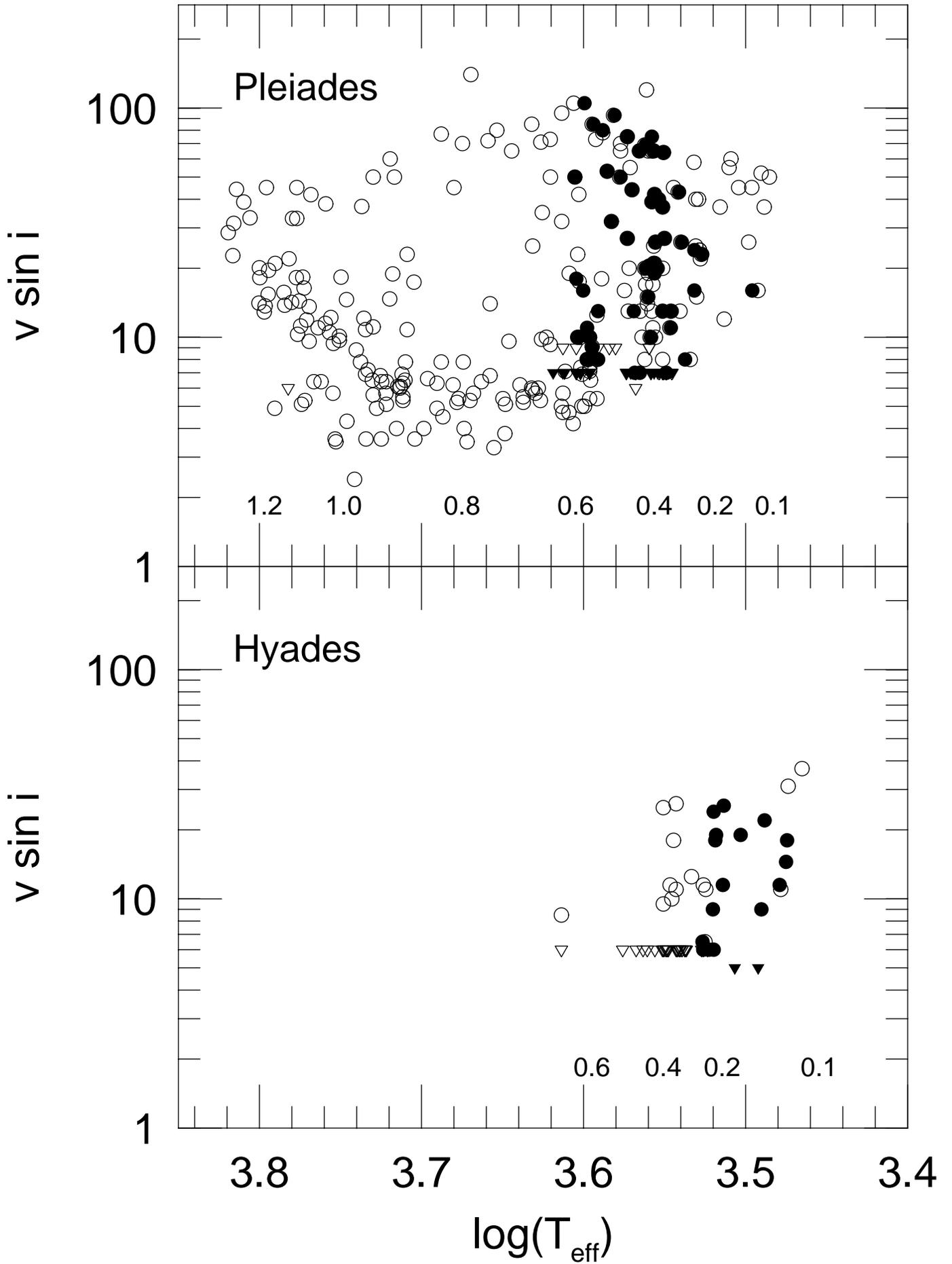
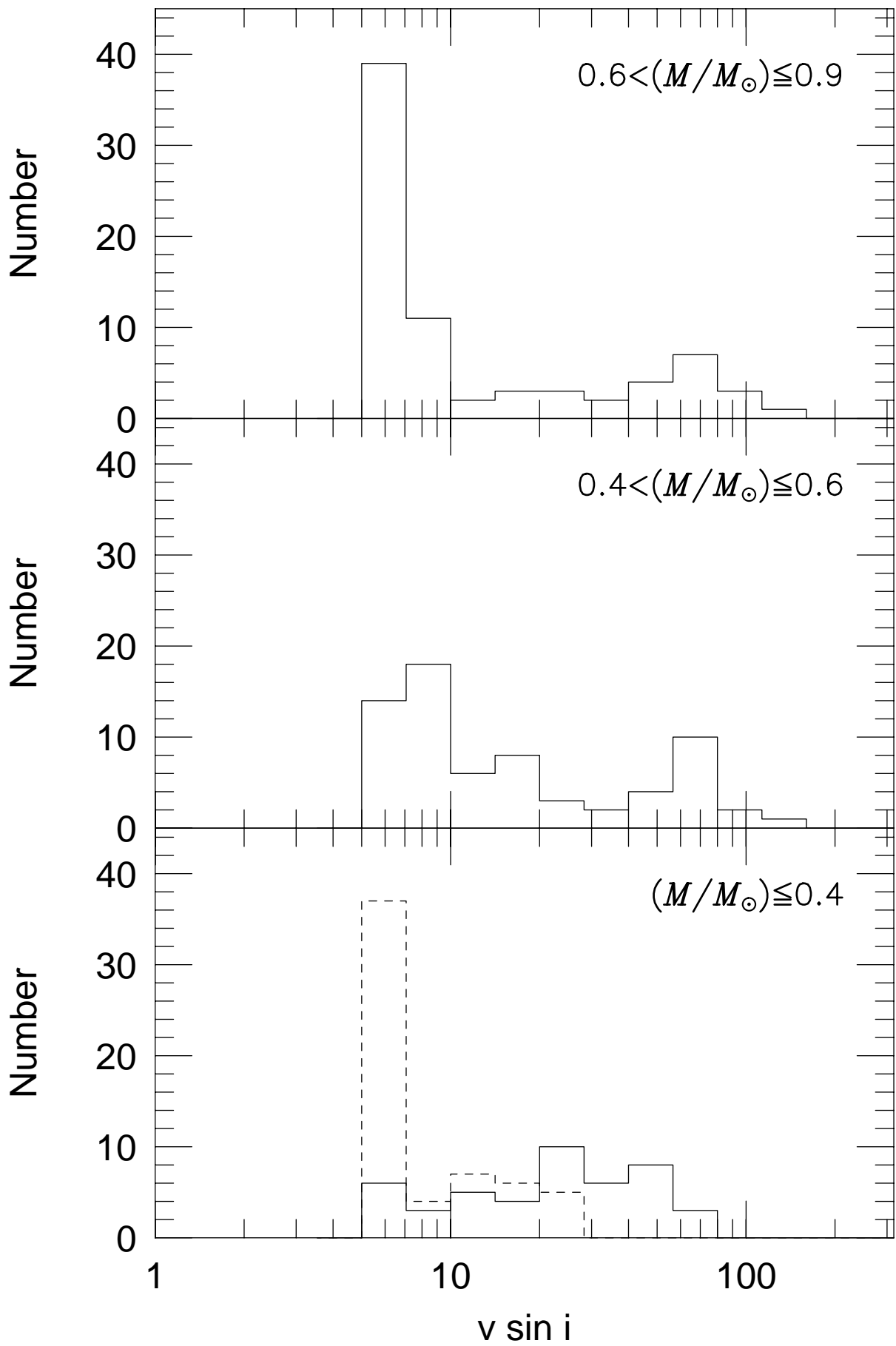
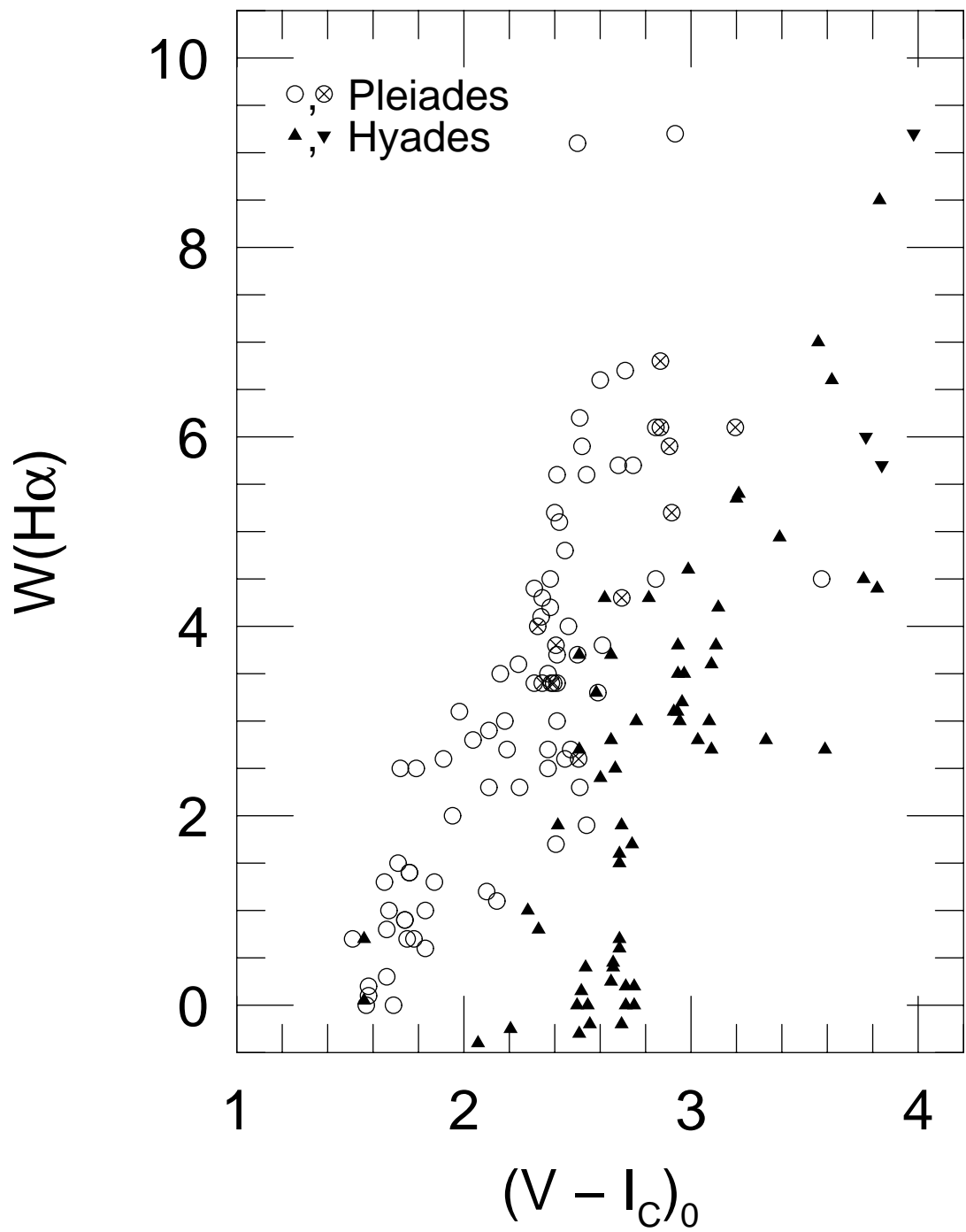
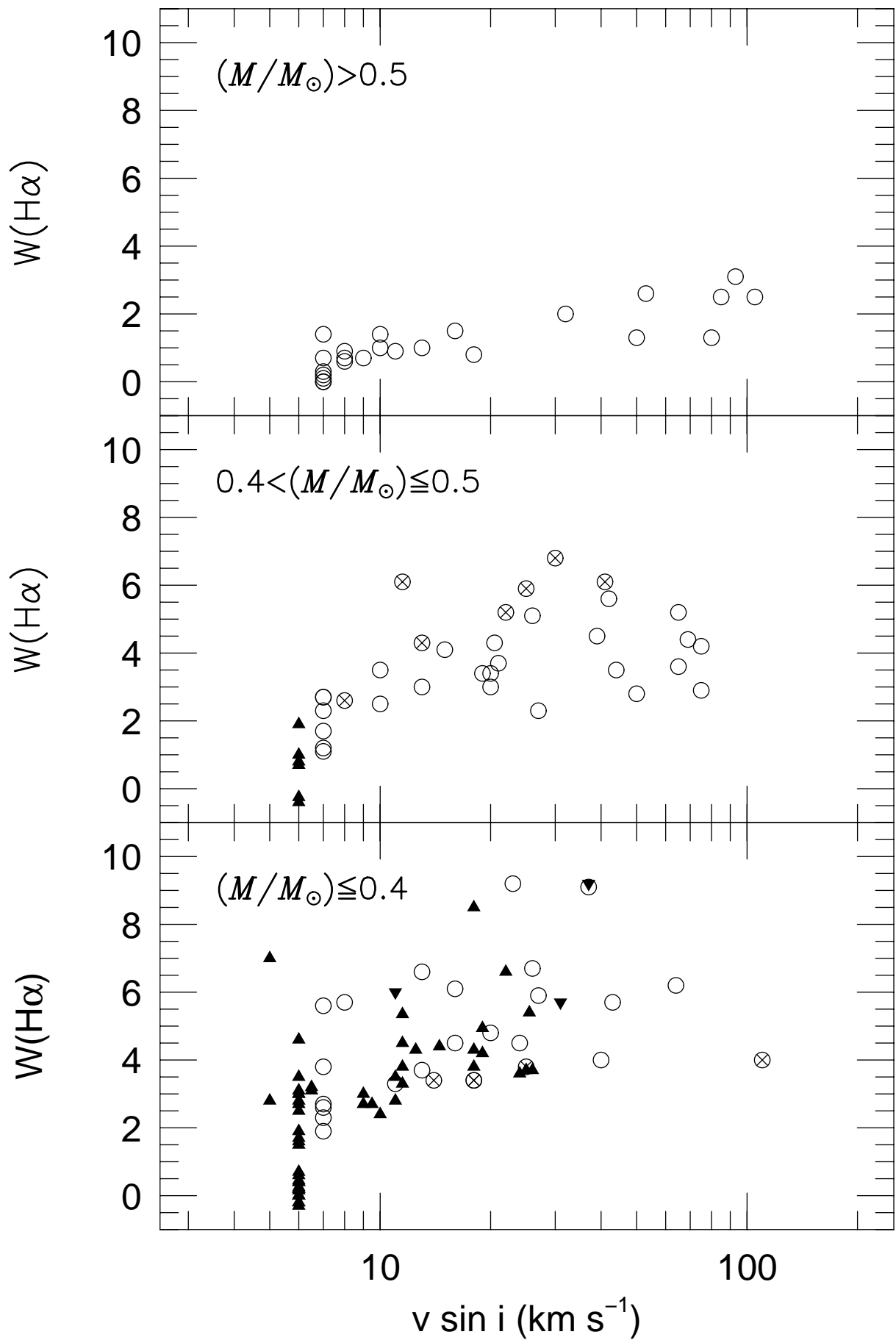
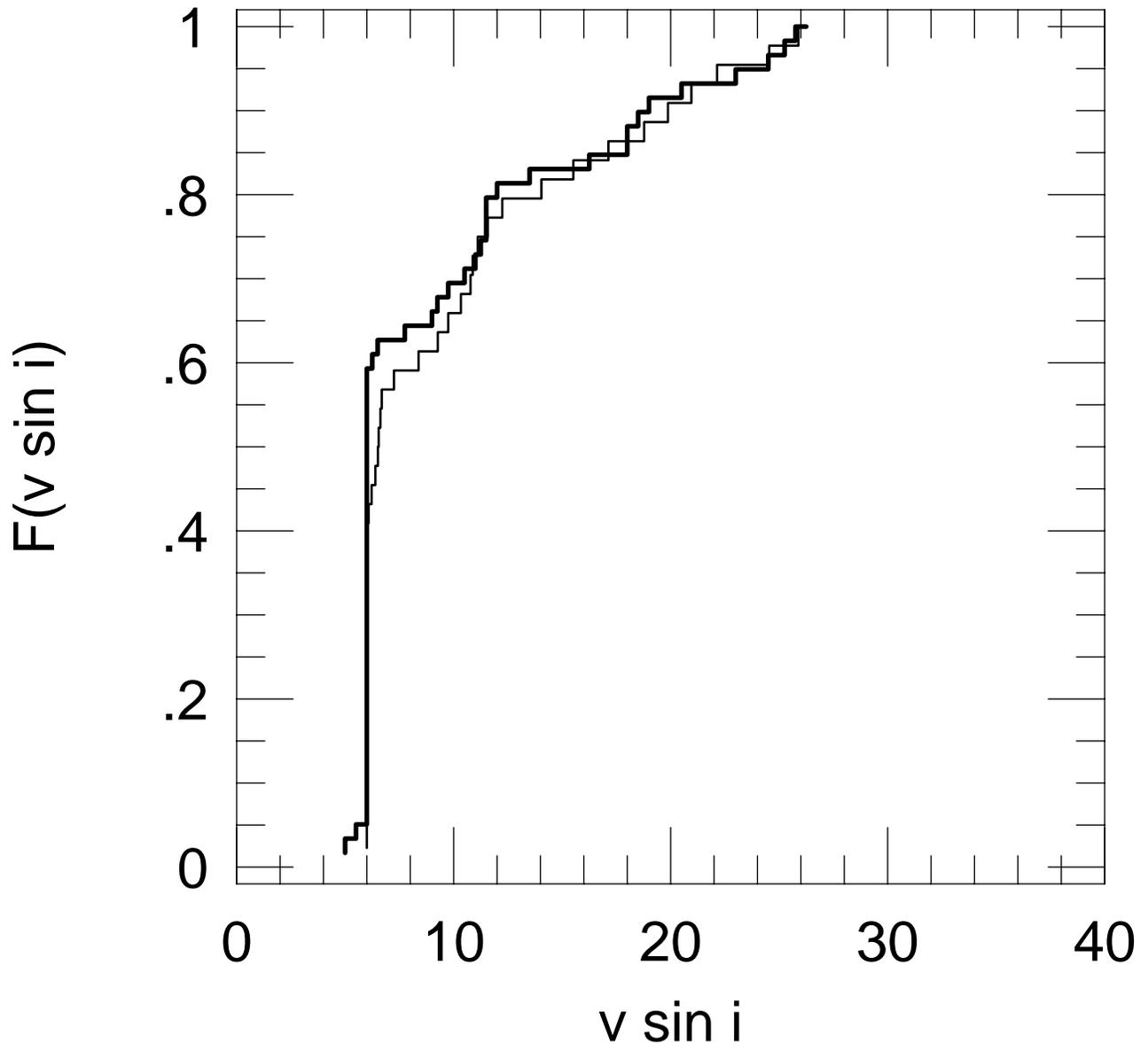
\end{document}